\documentclass[final,1pt,times]{elsarticle}

\usepackage{amssymb}
\usepackage{amsthm}

\usepackage{graphicx}
\usepackage{lineno}
\usepackage{colortbl}

\journal{Astroparticle Physics}

\begin{document}

\begin{frontmatter}

\title{Search for Relativistic Magnetic Monopoles with the ANTARES Neutrino Telescope}

\author[UPV]{S.~Adri\'an-Mart\'inez}
\author[IFIC]{J.A.~Aguilar}
\author[CPPM]{I.~Al~Samarai}
\author[Colmar]{A.~Albert}
\author[UPC]{M.~Andr\'e}
\author[Genova]{M.~Anghinolfi}
\author[Erlangen]{G.~Anton}
\author[IRFU/SEDI]{S.~Anvar}
\author[UPV]{M.~Ardid}
\author[NIKHEF]{A.C.~Assis Jesus}
% ...

\cortext[cor:1]{Corresponding author}
\fntext[tag:1]{Also at University of Leiden, the Netherlands}
\fntext[tag:2]{On leave at DESY, Platanenallee 6, D-15738 Zeuthen, Germany}
\fntext[tag:3]{Now at University of California - Irvine, 92697, CA, USA}
\fntext[tag:4]{Deceased}

\nopagebreak[3]
\address[UPV]{{Institut d'Investigaci\'o per a la Gesti\'o Integrada de les Zones Costaneres (IGIC) - Universitat Polit\`ecnica de Val\`encia. C/  Paranimf 1 , 46730 Gandia, Spain.}}\vspace*{0.15cm}
\address[IFIC]{{IFIC - Instituto de F\'isica Corpuscular, Edificios Investigaci\'on de Paterna, CSIC - Universitat de Val\`encia, Apdo. de Correos 22085, 46071 Valencia, Spain}}\vspace*{0.15cm}
\address[CPPM]{{CPPM, Aix-Marseille Universit\'e, CNRS/IN2P3, Marseille, France}}\vspace*{0.15cm}
\address[Colmar]{{GRPHE - Institut universitaire de technologie de Colmar, 34 rue du Grillenbreit BP 50568 - 68008 Colmar, France }}\vspace*{0.15cm}
\address[UPC]{{Technical University of Catalonia, Laboratory of Applied Bioacoustics, Rambla Exposici\'o, 08800 Vilanova i la Geltr\'u, Barcelona, Spain}}\vspace*{0.15cm}
\address[Genova]{{INFN - Sezione di Genova, Via Dodecaneso 33, 16146 Genova, Italy}}
\address[Erlangen]{{Friedrich-Alexander-Universit\"at Erlangen-N\"urnberg, Erlangen Centre for Astroparticle Physics, \newline
Erwin-Rommel-Str. 1, 91058 Erlangen, Germany}}\vspace*{0.15cm}
\address[IRFU/SEDI]{{Direction des Sciences de la Mati\`ere - Institut de recherche sur les lois fondamentales de l'Univers - Service d'Electronique des D\'etecteurs et d'Informatique, CEA Saclay, 91191 Gif-sur-Yvette Cedex, France}}\vspace*{0.15cm}
\address[NIKHEF]{{Nikhef, Science Park,  Amsterdam, The Netherlands}}\vspace*{0.15cm}
\address[APC]{{APC - Laboratoire AstroParticule et Cosmologie, UMR 7164 (CNRS, Universit\'e Paris 7 Diderot, CEA, Observatoire de Paris) 10, rue Alice Domon et L\'eonie Duquet 75205 Paris Cedex 13,  France}}
\address[LAM]{{LAM - Laboratoire d'Astrophysique de Marseille, P\^ole de l'\'Etoile Site de Ch\^ateau-Gombert, rue Fr\'ed\'eric Joliot-Curie 38,  13388 Marseille Cedex 13, France }}\vspace*{0.15cm}
\address[Bologna]{{INFN - Sezione di Bologna, Viale C. Berti-Pichat 6/2, 40127 Bologna, Italy}}
\address[Bologna-UNI]{{Dipartimento di Fisica dell'Universit\`a, Viale Berti Pichat 6/2, 40127 Bologna, Italy}}\vspace*{0.15cm}
\address[Pisa]{{INFN - Sezione di Pisa, Largo B. Pontecorvo 3, 56127 Pisa, Italy}}\vspace*{0.15cm}
\address[Roma]{{INFN - Sezione di Roma, P.le Aldo Moro 2, 00185 Roma, Italy}}\vspace*{0.15cm}
\address[Roma-UNI]{{Dipartimento di Fisica dell'Universit\`a La Sapienza, P.le Aldo Moro 2, 00185 Roma, Italy}}
\address[Clermont-Ferrand]{{Clermont Universit\'e, Universit\'e Blaise Pascal, CNRS/IN2P3, Laboratoire de Physique Corpusculaire, BP 10448, 63000 Clermont-Ferrand, France}}\vspace*{0.15cm}
\address[GEOAZUR]{{G\'eoazur - Universit\'e de Nice Sophia-Antipolis, CNRS/INSU, IRD, Observatoire de la C\^ote d'Azur and Universit\'e Pierre et Marie Curie, BP 48, 06235 Villefranche-sur-mer, France}}\vspace*{0.15cm}
\address[Bari]{{INFN - Sezione di Bari, Via E. Orabona 4, 70126 Bari, Italy}}\vspace*{0.15cm}
\address[COM]{{COM - Centre d'Oc\'eanologie de Marseille, CNRS/INSU et Universit\'e de la M\'editerran\'ee, 163 Avenue de Luminy, Case 901, 13288 Marseille Cedex 9, France}}\vspace*{0.15cm}
\address[LNS]{{INFN - Laboratori Nazionali del Sud (LNS), Via S. Sofia 62, 95123 Catania, Italy}}\vspace*{0.15cm}
\address[UPS]{{Univ Paris-Sud , 91405 Orsay Cedex, France}}\vspace*{0.15cm}
\address[KVI]{{Kernfysisch Versneller Instituut (KVI), University of Groningen, Zernikelaan 25, 9747 AA Groningen, The Netherlands}}\vspace*{0.15cm}
\address[Pisa-UNI]{{Dipartimento di Fisica dell'Universit\`a, Largo B. Pontecorvo 3, 56127 Pisa, Italy}}\vspace*{0.15cm}
\address[NIOZ]{{Royal Netherlands Institute for Sea Research (NIOZ), Landsdiep 4,1797 SZ 't Horntje (Texel), The Netherlands}}\vspace*{0.15cm}
\address[Bamberg]{{Dr. Remeis-Sternwarte and ECAP, Universit\"at Erlangen-N\"urnberg,  Sternwartstr. 7, 96049 Bamberg, Germany}}\vspace*{0.15cm}
\address[UU]{{Universiteit Utrecht, Faculteit Betawetenschappen, Princetonplein 5, 3584 CC Utrecht, The Netherlands}}\address[UvA]{{Universiteit van Amsterdam, Instituut voor Hoge-Energie Fysika, Science Park 105, 1098 XG Amsterdam, The Netherlands}}\vspace*{0.15cm}
\address[MSU]{{Moscow State University,Skobeltsyn Institute of Nuclear Physics,Leninskie gory, 119991 Moscow, Russia}}\address[Catania]{{INFN - Sezione di Catania, Viale Andrea Doria 6, 95125 Catania, Italy}}\vspace*{0.15cm}
\address[Catania-UNI]{{Dipartimento di Fisica ed Astronomia dell'Universit\`a, Viale Andrea Doria 6, 95125 Catania, Italy}}\address[IRFU/SPP]{{Direction des Sciences de la Mati\`ere - Institut de recherche sur les lois fondamentales de l'Univers - Service de Physique des Particules, CEA Saclay, 91191 Gif-sur-Yvette Cedex, France}}\vspace*{0.15cm}
\address[WIN]{{University of Wisconsin - Madison, 53715, WI, USA}}\vspace*{0.15cm}
\address[Livorno]{{Accademia Navale di Livorno, Livorno, Italy}}\vspace*{0.15cm}
\address[ISS]{{Institute for Space Sciences, R-77125 Bucharest, M\u{a}gurele, Romania}}\vspace*{0.15cm}
\address[IPHC]{{IPHC-Institut Pluridisciplinaire Hubert Curien - Universit\'e de Strasbourg et CNRS/IN2P3  23 rue du Loess, BP 28,  67037 Strasbourg Cedex 2, France}}\vspace*{0.15cm}
\address[ITEP]{{ITEP - Institute for Theoretical and Experimental Physics, B. Cheremushkinskaya 25, 117218 Moscow, Russia}}\vspace*{0.15cm}
\address[Genova-UNI]{{Dipartimento di Fisica dell'Universit\`a, Via Dodecaneso 33, 16146 Genova, Italy}}\vspace*{0.15cm}

\begin{abstract}
Magnetic monopoles are predicted in various unified gauge models and could be produced at intermediate mass scales. Their detection in a neutrino telescope is facilitated by the large amount of light emitted compared to that from muons. This paper reports on a search for upgoing relativistic magnetic monopoles with the ANTARES neutrino telescope using a data set of 116 days of live time taken from December 2007 to December 2008. The one observed event is consistent with the expected atmospheric neutrino and muon background, leading to a 90\% C.L. upper limit on the monopole flux between $1.3\times10^{-17}$ and $8.9\times10^{-17}$ cm$^{-2}$s$^{-1}$sr$^{-1}$ for monopoles with velocity $\beta \geq0.625$.
\end{abstract}


\end{frontmatter}

%\begin{linenumbers}

%%%%%%%
%INTRODUCTION
\section{Introduction}

Magnetic monopoles are hypothetical particles first proposed by Pierre Curie in 1894~\cite{ref:Curie}. In 1931 Dirac demonstrated that the existence of magnetic monopoles naturally leads to and thus explains the quantization of the electric charge~\cite{ref:Dirac}. In 1974, 't~Hooft and Polyakov discovered independently that, in certain spontaneously broken gauge theories, magnetic monopoles are not only a possibility, but a requirement~\cite{ref:thooft,ref:polyakov}.

Despite intensive search efforts,  no particles possessing magnetic charge have been detected up to now~\cite{ref:pdg}, although some reports of magnetic monopole detections have been claimed~\cite{ref:price, ref:cabrera, ref:caplin} and subsequently withdrawn~\cite{ref:price2, ref:huber}. Nevertheless, direct searches for monopoles in cosmic rays over the past decade have provided stringent flux limits below the original Parker bound $\Phi_P \sim 10^{-15}$~cm$^{-2}$s$^{-1}$sr$^{-1}$~\cite{ref:Parker} (cf. Sec.~\ref{Sec_MM_th}). The MACRO experiment obtained a flux upper limit at the level of $1.4 \times 10^{-16}$~cm$^{-2}$s$^{-1}$sr$^{-1}$ for the monopole velocity range $4\times10^{-5}<\beta<1$~\cite{ref:MACRO}. For relativistic magnetic monopoles ($\beta\gtrsim0.8$), stronger bounds were reported by the Baikal and AMANDA neutrino telescopes. The Baikal neutrino telescope NT200 has set an upper limit of $4.6 \times 10^{-17}$~cm$^{-2}$s$^{-1}$sr$^{-1}$ on the flux of monopoles for $\beta \simeq 1$~\cite{ref:BAIKAL}, while the AMANDA-II limit at $\beta \simeq 1$ is $3.8 \times 10^{-17}$~cm$^{-2}$s$^{-1}$sr$^{-1}$~\cite{ref:AMANDA}. More recently, experiments based on radio detection reported stronger upper limits for ultra-relativistic magnetic monopoles. The RICE experiment at the South Pole obtained an upper limit of $10^{-18}$~cm$^{-2}$s$^{-1}$sr$^{-1}$ for Lorentz boost factors $10^7\le\gamma\le10^{12}$~\cite{ref:RICE} and ANITA improved the flux limit to $10^{-19}$~cm$^{-2}$s$^{-1}$sr$^{-1}$ for $10^{10}\le\gamma\le10^{13}$~\cite{ref:ANITA}.

In this paper, a search for upgoing relativistic magnetic monopoles is presented for one year of data taking with the ANTARES detector~\cite{ref:AntaresDet}. The outline of this paper is as follows: Section~\ref{Sec_MM} introduces the magnetic monopole theory and the expected signal from such particles crossing a neutrino telescope. The ANTARES underwater neutrino telescope is briefly presented in Section~\ref{Sec_Det}. Section~\ref{Sec_Sim} describes the simulation and the reconstruction algorithm. Finally the search strategy is discussed in Section~\ref{Sec_Ana} and results are shown in Section~\ref{Sec_Res}.

%%%%%%%
%SECTION 1
\section{Magnetic monopoles and their signal in a neutrino telescope}\label{Sec_MM}
\subsection{Magnetic monopoles}\label{Sec_MM_th}

In his paper of 1931, Dirac introduced magnetic monopoles in quantum theory and showed that the existence of a particle carrying a magnetic pole leads to the quantization of electric charge~\cite{ref:Dirac}. In addition, the magnetic charges of the hypothetical monopoles must be quantized according to the Dirac quantization condition $g=kg_D$, where $g_D=\frac{\hbar c}{2e}$ is the minimal magnetic charge, also called the ``Dirac charge'', $k$ is an integer and $e$ is the elementary electric charge.

In 1974, 't~Hooft and Polyakov independently discovered monopole solutions in Georgi-Glashow gauge theories based on the SO(3) group~\cite{ref:thooft,ref:polyakov}. It was then realized that any unification model in which the U(1) subgroup of electromagnetism is embedded in a semi-simple gauge group and which is spontaneously broken by the Higgs mechanism possesses monopole-like solutions. Monopoles appear as solitons that behave like particles in the classical theory, with a mass of the order of $M_{mon}\sim \alpha^{-1}\Lambda$, where $\alpha$ is the fine structure constant and $\Lambda$ is the unification mass scale or the intermediate mass scale of the underlying theory. Such magnetic monopoles are predicted in any Grand Unified Theory (GUT) in which a larger gauge group breaks down into a semi-simple subgroup containing the explicit U(1) group of electromagnetism. The mass predicted for the monopoles can range from $10^{4}$ GeV to $10^{20}$ GeV depending on the specific model~\cite{ref:Preskill}.

In such unified gauge models, magnetic monopoles would be produced in the early Universe by the Kibble mechanism~\cite{ref:Kibble}. In the case where monopoles were created at the GUT phase transition with a mass scale $\Lambda\sim 10^{15}$ GeV, the inflationary scenario strongly dilutes the monopole density, thereby solving the monopole overdensity problem~\cite{ref:Guth}. For monopoles produced in later phase transitions in the early Universe below $\Lambda \sim 10^{11}$ GeV, larger fluxes of monopoles with mass $10^{7} - 10^{13}$ GeV are expected~\cite{ref:Lazarides}. The latter scenario may produce an observable abundance of the so-called intermediate-mass monopoles if it occurs after cosmic inflation. Magnetic monopoles have been proposed as candidates for ultra-high energy cosmic rays provided that they do not catalyze nucleon decay~\cite{ref:Kephart}. Many other alternative scenarios have been put forward to solve the cosmological monopole problem and thus the hypothesis that monopoles are still present in the Universe is far from being dismissed. However, a stringent phenomenological upper bound is given by the Parker limit, $\Phi_P \sim 10^{-15}~$cm$^{-2}$ s$^{-1}$ sr$^{-1}$, which results from requiring the survival of the galactic magnetic field~\cite{ref:Parker}. A stronger bound derived from the survival of a small galactic seed field and known as the extended Parker bound is $\Phi_P \sim 1.2\times10^{-16} (M_{mon}/10^{17}GeV)\;$cm$^{-2}$ s$^{-1}$ sr$^{-1}$~\cite{ref:Adams}.

Measurements and estimates of cosmic magnetic fields suggest that magnetic monopoles lighter than $10^{14}$ GeV could have been accelerated to relativistic velocities, acquiring typical kinetic energies of $10^{15}$ GeV~\cite{ref:Wick}. Magnetic monopoles are expected to lose energy significantly when crossing the Earth~\cite{ref:Balestra} due to their large equivalent electric charge ($g_D\simeq68.5e$). Using a simple Earth model, calculations suggest, however, that magnetic monopoles with mass below $\sim 10^{14}$ GeV would be detectable in a neutrino telescope after crossing the Earth, despite their substantial energy loss~\cite{ref:Derkaoui}.

\subsection{Signal in a neutrino telescope}

\begin{figure}[h!]
   \begin{center}
      \includegraphics[height=2.5in]{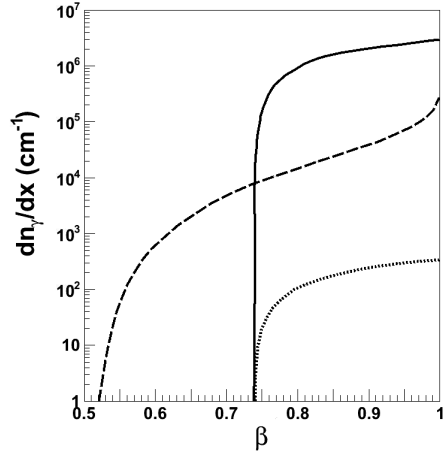}
   \end{center}
   \caption{\footnotesize Number of Cherenkov photons in the $300-600$ nm wavelength range emitted per cm in sea water from a monopole with $g=g_D$ (solid line) and from $\delta-$rays produced along its path (dashed line) as a function of the velocity $\beta$ of the monopole. For comparison, the direct Cherenkov emission from a muon is also shown (dotted line). The number of photons scales as the square of the monopole charge.}
   \label{deltaraychegraph}
\end{figure}

The detection of magnetic monopoles in a neutrino telescope is based on the same principle as the detection of high energy muons. Tompkins~\cite{ref:Tompkins} showed that, as for electric charges, magnetically charged particles produce Cherenkov emission when their velocity is higher than the Cherenkov threshold $\beta_{th}=1/n$, where $n$ is the phase refractive index of the medium. The number of photons emitted per unit length and wavelength, for $\beta>\beta_{th}$, can be expressed as:
\begin{equation}\label{eq1}
 \frac{d^2n_{\gamma}}{d\lambda dx}=\frac{2\pi \alpha}{\lambda^2} \left(\frac{ng}{e}\right)^2 \left(1-\frac{1}{n^2\beta^2}\right)
\end{equation}
where $n_{\gamma}$ is the number of emitted photons and $\lambda$ their wavelength. For the refractive index of sea water ($n\sim1.35$), fast monopoles with $g=g_D$ are expected to emit about $8550$ times more Cherenkov photons than muons of the same velocity. Moreover, below the Cherenkov threshold $\beta_{th}=0.74$, a magnetic monopole of velocity $\beta\gtrsim 0.51$ ionizes sea water leading to indirect Cherenkov emission from knock off electrons ($\delta-$rays) produced along its path~\cite{ref:vanRens}. Figure~\ref{deltaraychegraph} shows the number of photons emitted per centimetre of track length as a function of the velocity of the incoming monopole for the wavelength range relevant in neutrino telescopes (300-600 nm). The same quantity is also shown for a relativistic muon. Contributions from radio-luminescence of water, pair production and Bremsstrahlung induced by magnetic monopoles are negligible compared to the direct and indirect Cherenkov light and are not taken into account in this analysis.

%%%%%%%
%SECTION 2
\section{The ANTARES neutrino telescope}\label{Sec_Det}

The ANTARES detector is an underwater telescope immersed in the Western Mediterranean Sea at a depth of 2475~m~\cite{ref:AntaresDet}. In the final configuration the detector consists of 885 optical modules (glass spheres housing a photomultiplier) on twelve mooring lines. Each detector line comprises 25 storeys, each of them housing three optical modules, and is connected via an interlink cable to a Junction Box. The Junction Box itself is connected to the shore station at La Seyne-sur-Mer by a 42 km long electro-optical cable. During the construction of the detector, data were taken with a different number of lines in different periods. 

The analysis has been performed using data taken from December 2007 to December 2008. Quality requirements are applied to select data from periods with low levels of bioluminescent activity and a well calibrated detector. After this selection the data is equivalent to a total of 136 days of live time: 43 days with 12 lines, 46 days with 10 lines and 47 days with 9 lines.

%%%%%%%
%SECTION 3
\section{Simulation and reconstruction}\label{Sec_Sim}

\subsection{Monte Carlo simulations}

\begin{figure}[h!]
   \begin{center}
      \includegraphics[height=2.5in]{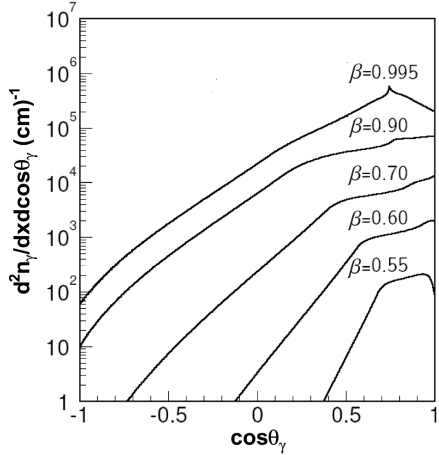}
   \end{center}
   \caption{\footnotesize{Angular distributions of Cherenkov photons from $\delta-$rays produced by a monopole with $g=g_D$ as a function of the emission angle $\theta_\gamma$ between the photon and the monopole. Distributions are shown for monopole velocities $\beta = 0.55$, 0.60, 0.70, 0.90 and 0.995. }}
  \label{Dispersionangulaire}
\end{figure}

Upgoing magnetic monopoles with one unit of Dirac charge ($g=g_D$) have been simulated using a Monte Carlo program based on GEANT3 \cite{ref:GEANT} for ten ranges of velocities in the region $\beta=[0.550,0.995]$, with a flat distribution inside each bin. The simulation is independent of the magnetic monopoles mass and the incoming direction of monopoles was distributed isotropically over the lower hemisphere. The number of direct Cherenkov photons emitted by the magnetic monopoles is computed using Eq. (\ref{eq1}), with an emission angle with respect to the monopole direction defined by $cos(\theta_{\gamma})=1/\beta n$. For photons emitted from $\delta-$rays, the angular dispersion is calculated numerically~\cite{ref:Ricol} from the multiple scattering of electrons in water \cite{ref:Standard}. Figure~\ref{Dispersionangulaire} shows the angular distribution of Cherenkov photons  from $\delta-$rays with respect to the monopole direction for several values of the monopole velocity. 

The simulation of emitted photons is processed inside a cylindrical volume surrounding the instrumented volume. A radius of $480$~m (eight times the absorption length),  four times larger than that used for the standard ANTARES muon simulation, is chosen in order to take into account the large amount of light emitted by a magnetic monopole.

When searching for upgoing magnetic monopoles, the main source of background events is due to upgoing muons induced by atmospheric neutrinos and downgoing atmospheric muons wrongly reconstructed as upgoing. The simulation of downgoing atmospheric muons produced by the interaction of cosmic rays with nuclei in the upper atmosphere is carried out using the CORSIKA air shower program~\cite{ref:Corsika} in combination with the QGSJET code for the description of hadronic interactions~\cite{ref:QGSJET}. The cosmic ray spectrum is simulated according to the H\"orandel model~\cite{ref:Horandel}. Upgoing atmospheric neutrinos from the decay of pions and kaons are generated assuming the Bartol atmospheric neutrino flux model~\cite{ref:Bartol, ref:Bartol2} and are combined with neutrinos coming from the decay of charm mesons as produced by the RQPM model \cite{ref:RQPM}.

In order to match the real detector conditions, the simulations are performed separately for each detector configuration (9, 10 and 12 lines) with the dead channel mapping and optical background rates corresponding to each data taking period.

\subsection{Trigger selection}

The ANTARES data acquisition is based on the "all-data-to-shore" concept~\cite{ref:AntaresDAQ}. In this scheme, all photomultiplier signals above a given threshold equivalent to $\sim0.3$ photo-electrons are transmitted to shore, where they are filtered using different trigger conditions and finally transferred to mass storage.

For both the data and Monte Carlo simulation samples, pattern recognition of the fired channels (hits) is performed within a time window of 2.2 $\mu$s, the typical time that a relativistic particle with $\beta=0.5$ takes to cross the detector. Two standard pattern recognition templates based on local coincidences have been applied. A local coincidence is defined either as an occurrence of two hits on two separate optical modules of a single storey within 20 ns, or one single hit of large amplitude, typically more than 3 photoelectrons. The first trigger (directional trigger) requires five local coincidences within the triggering time window anywhere in the detector that are causally connected. The second trigger (cluster trigger) requires two so-called T3-clusters within the trigger window. A T3-cluster is a combination of two local coincidences in adjacent or next-to-adjacent storeys within 100 ns or 200 ns, respectively. When an event is triggered, the amplitude and time of all hits are recorded during a window from $2.2\;\mu$s before the first hit participating in the trigger until  $2.2\;\mu$s after the last hit of the trigger. More than 85 \% of monopoles with a velocity $\beta > 0.58$ that produce at least six detected hits pass the above trigger conditions.

\subsection{Reconstruction algorithm}\label{ss_rec}

The standard track reconstruction assumes that particles travel at the speed of light. In order to improve the sensitivity for magnetic monopoles travelling with lower velocities, one of the ANTARES tracking algorithms~\cite{ref:AntaresBBFit} has been modified so as to leave the velocity as a free parameter to be determined by the track fit. This algorithm, based on the minimization of time residuals using the least square method (see below), has a very stringent hit selection, which leads to a robust reconstruction with respect to background hits. 

The modified algorithm performs two independent fits: a track fit and a bright point fit. The former reconstructs the track of a particle crossing the detector at a velocity $\beta_{free}$, introduced as a free parameter, while the latter reconstructs the event as a point-like light source. Both fits minimize the same quality function Q defined as~:
\begin{equation}
Q=\sum_{i=1}^{N}\left[\frac{(t_\gamma-t_i)^2}{\sigma_i^2}+A_i\right]
\end{equation}
The first term on the right hand side is the sum for $N$ hits of the square of the time residuals, where $t_\gamma$ is the expected time of a hit, $t_i$ the measured time and $\sigma_i$ the estimated time uncertainty for hit $i$. The second term $A_i\sim q_i d_i$ is introduced to penalize hits with large charge $q_i$ combined with a large distance of closest approach $d_i$ between the track and the detection line.

\begin{figure}[h!]
   \begin{center}
      \includegraphics[height=1.6in]{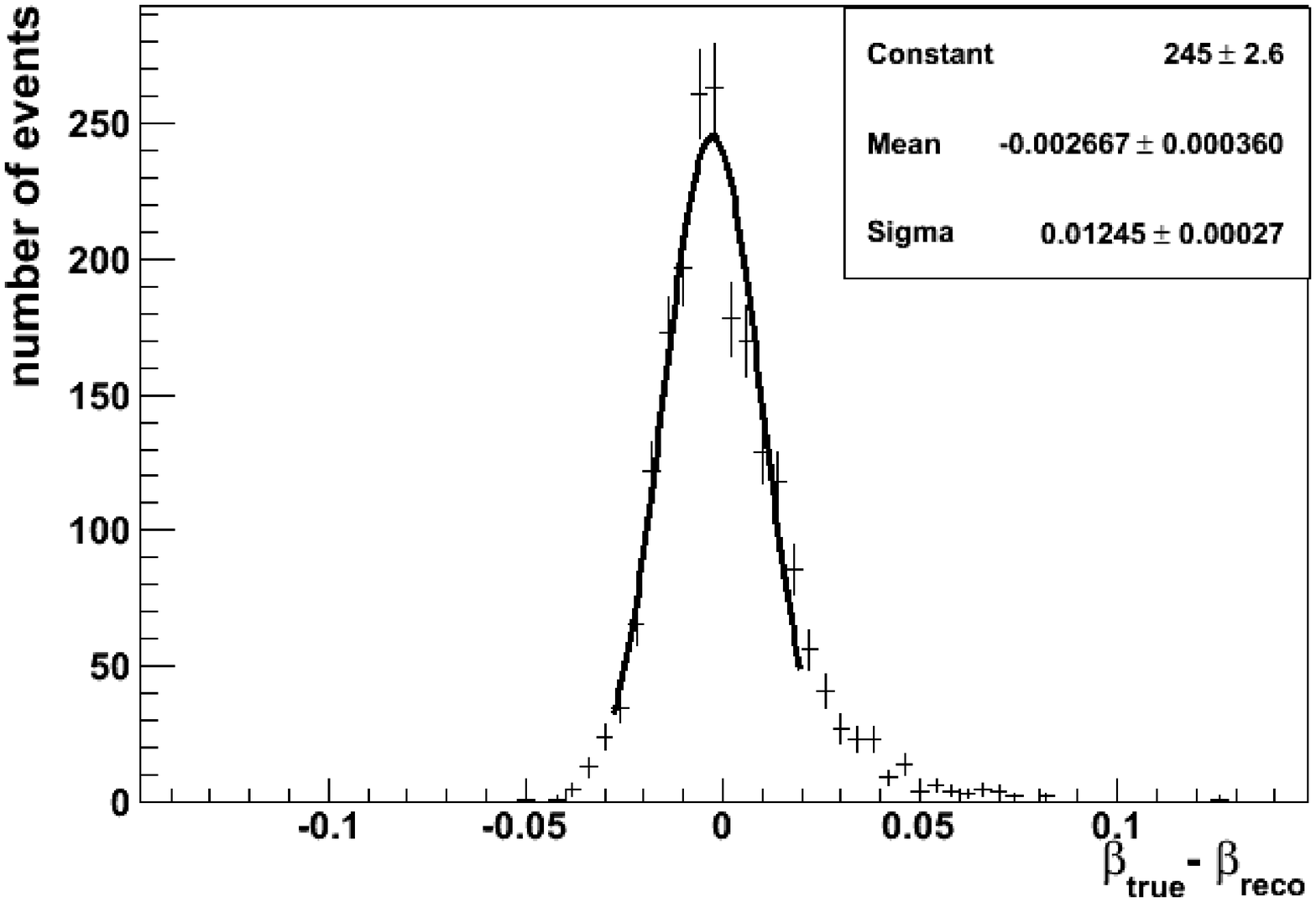}
      \includegraphics[height=1.6in]{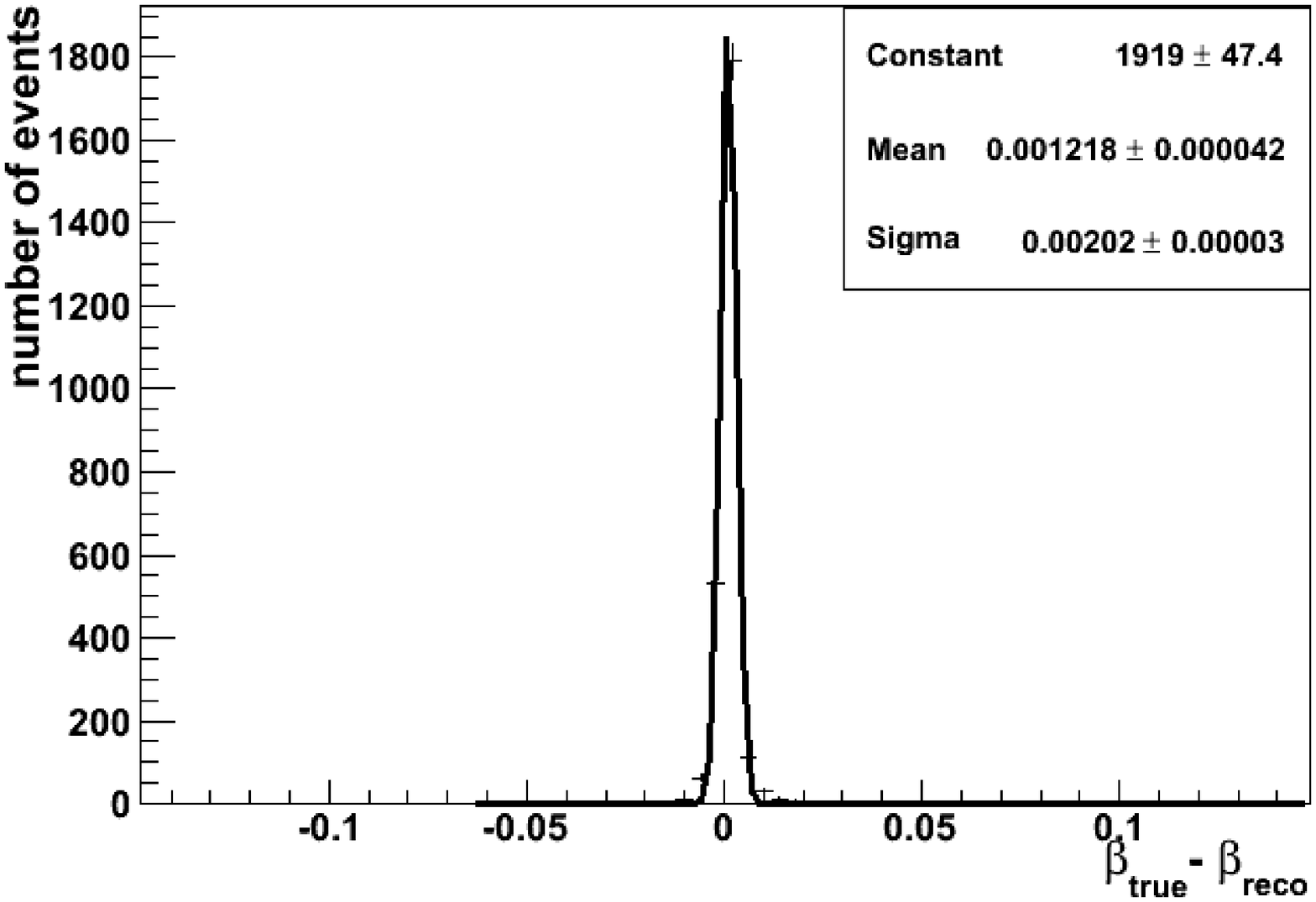}
   \end{center}
   \caption{Resolution on the reconstructed velocity for monopoles generated at $\beta$ = [0.625,0.675] (left panel) and $\beta$= [0.825,0.875] (right panel) after the final selection cuts described in Sec.~\ref{Sec_Ana} are applied.}
   \label{fig_res}
\end{figure}

After imposing the final selection cuts described in Sec.~\ref{Sec_Ana}, the modified tracking algorithm yields to an approximately Gaussian resolution $\sigma_\beta$ on the magnetic monopole reconstructed velocity of about $\sigma_\beta \simeq 0.025$ for velocities lower than the Cherenkov threshold. The resolution improves to $\sigma_\beta \simeq 0.003$ for higher velocities. The speed resolution is shown in Fig.~\ref{fig_res} for magnetic monopoles simulated in two different velocity ranges.

%%%%%%%
%SECTION4
\section{Event selection}\label{Sec_Ana}

The search strategy is based on a blind analysis in order to avoid any experimental bias~\cite{ref:Roodman}. Both the first level selection cuts and the final event selection based on the optimization of the Model Discovery Factor (MDF)~\cite{ref:Punzi,ref:Hill} were tuned on Monte Carlo simulated samples. The background simulations have been verified after each step of the event selection with a test sample of 15\% of the selected data, equivalent to 20 days out of the total 136 days of live time.

\subsection{First level selection}\label{ss_prel}

\begin{figure}[h!]
   \begin{center}
      \includegraphics[height=2.5in]{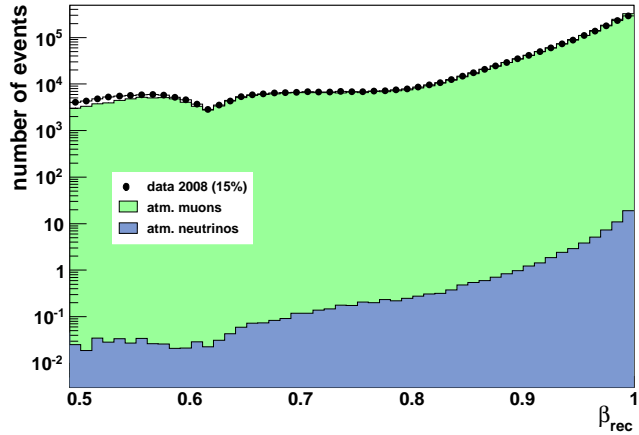}
   \end{center}
   \caption{Distributions of atmospheric muons and upgoing neutrino-induced muons compared with the 15\% data sample as a function of the reconstructed velocity $\beta_{rec}$. The distribution of simulated atmospheric muons has been scaled with a normalization factor of 1.8.}
   \label{betadist2}
\end{figure}

Both Monte Carlo events and data (the sample of 15\%) have been reconstructed with the modified algorithm introduced in Sec.~\ref{ss_rec}. The data were compared to the Monte Carlo expectation over the full range of reconstructed track velocities, as shown in Fig.~\ref{betadist2}. A normalization factor of 1.8 has been applied to the simulated sample of atmospheric muons, which is consistent with the expected uncertainties on the optical module angular acceptance of Cherenkov light from downgoing particles and on parameters of the atmospheric muon flux model, such as the primary cosmic ray composition and the hadronic interaction models. Good agreement between the simulation and experimental distributions is observed in the region $\beta_{rec}>\;$0.6, while the simulation underestimates the data for lower velocities. Monte Carlo studies indicate that the region below $\beta=\;$0.625 is very sensitive to the time dependence of the optical background, consequently events in this region are not considered for further analysis.

\begin{figure}[h!]
   \begin{center}
      \includegraphics[height=2.5in]{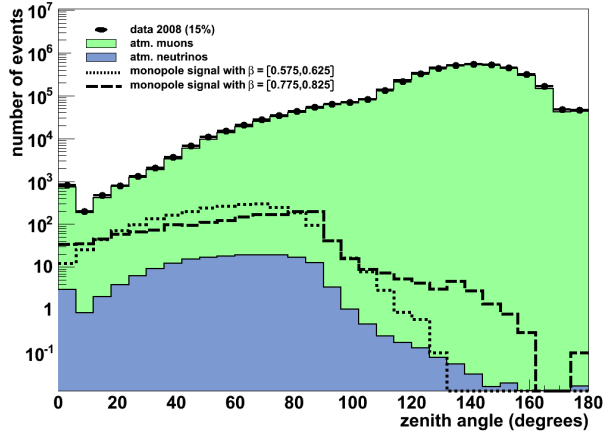}               
   \end{center}
   \caption{Zenith angle distributions of simulated downgoing atmospheric muons (with the normalization factor applied) and simulated muons induced by upgoing atmospheric neutrinos compared with the 15\% data sample. The distributions of simulated magnetic monopoles generated at $\beta=[0.575,0.625]$ (dotted line) and $\beta=[0.775,0.825]$ (dashed line) are also shown with arbitrary absolute normalization. Only tracks reconstructed using hits on at least two lines are considered in this plot.}
   \label{Zenith}
\end{figure}

Figure~\ref{Zenith} shows the zenith angle distribution of the reconstructed tracks for simulated atmospheric muons and muons induced by upgoing atmospheric neutrinos compared with the data sample. Upgoing magnetic monopoles generated in two different velocity regions are also presented. 

Due to the large background from atmospheric muons in the downgoing direction, only upgoing tracks with a zenith angle smaller than $90^{\circ}$ are selected. To reduce the number of poorly reconstructed events, only those for which the track is reconstructed using hits from at least two lines are kept. The third first level selection cut rejects events that fit the bright point hypothesis better than the track hypothesis, cf. Sec.~\ref{ss_rec}. This suppresses events dominated by electromagnetic and hadronic showers and reduces by a large fraction the number of misreconstructed atmospheric muons.

\subsection{Final selection}\label{ss_final}
\begin{figure}[h!]
   \begin{center}

      \includegraphics[height=2.5in]{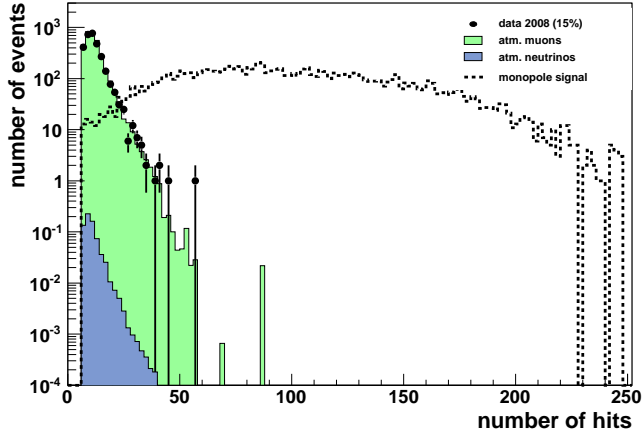}
 
   \end{center}
   \caption{Distributions of the number of hits used in the track reconstruction for events passing the first level selection cuts. The solid histograms correspond to the simulated downgoing atmospheric muons (normalization factor applied) and the muons from upgoing atmospheric neutrinos. The points correspond to the 15\% sample of data. The dashed line indicates the distribution for magnetic monopoles generated in the range $\beta=[0.775,0.825]$. All these distributions are shown for events for which the reconstructed velocity is restricted to $\beta_{rec}=[0.775,0.825]$. }
   \label{nhitdist}
\end{figure}

The final event selection was performed by optimizing the Model Discovery Factor by cutting on discriminating variables. The first discriminating variable is the number of hits associated to the track used by the reconstruction algorithm. The large amount of light emitted by a monopole compared to that of atmospheric muons or muons induced by atmospheric neutrinos makes this a particularly powerful discriminant. This is demonstrated in Fig.~\ref{nhitdist} where the distributions of the numbers of hits are shown for the simulated atmospheric events and for magnetic monopoles simulated in the range $\beta=[0.775,0.825]$. For all tracks in these distributions the reconstructed velocity is restricted to $\beta_{rec}=[0.775,0.825]$. 

\begin{figure}[h!]
   \begin{center}

      \includegraphics[height=2.5in]{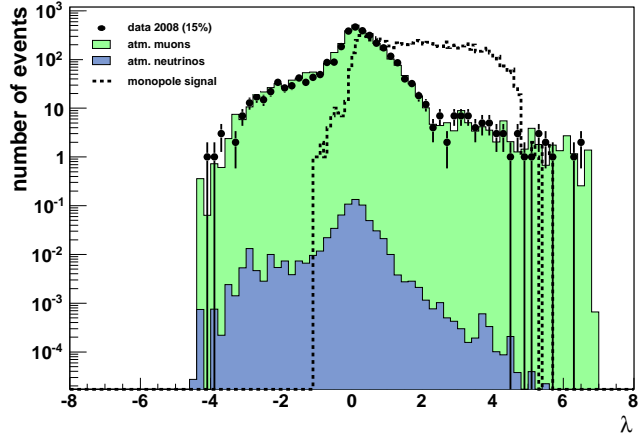}
 
   \end{center}
   \caption{Distributions of the parameter $\lambda$ for events passing the first level selection cuts. The solid histograms correspond to the simulated downgoing atmospheric muons (normalization factor applied) and muons from upgoing atmospheric neutrinos. The points correspond to the 15\% sample of data. The dashed line are magnetic monopoles generated in the range $\beta=[0.775,0.825]$ drawn with an arbitrary normalization. All these distributions are shown for events for which the reconstructed velocity is restricted to $\beta_{rec}=[0.775,0.825]$.}
   \label{lambdadist}
\end{figure}

A second discriminating variable was introduced in order to further reduce the background, in particular for velocities below the Cherenkov threshold where the light emission is less. Two different track-reconstruction fits for each event are performed. In the first fit, the velocity $\beta_{rec}$ is fixed at 1, whereas the second modified algorithm allows $\beta_{rec}$ as a free parameter in the fit procedure. The discriminating parameter $\lambda$ is then defined as

\begin{equation}
 \lambda=\log\left(\frac{Q_t({\beta_{rec}=1})}{Q_t({\beta_{rec}=free})}\right),
\end{equation}
where $Q_t({\beta_{rec}=1})$ and $Q_t({\beta_{rec}=free})$ are the track quality parameters for fixed and free $\beta_{rec}$, respectively. With this definition, it is expected that $\lambda$ is positive for monopoles and negative for atmospheric events. This feature is confirmed in Fig.~\ref{lambdadist}, where distributions of $\lambda$ are displayed for events reconstructed in the range $\beta_{rec}=[0.775,0.825]$. 
 
The selection cuts were optimized by minimizing the MDF for a $5\sigma$ discovery at $90\%$ probability. This minimization was performed by varying the cuts on the number of hits and the $\lambda$ parameter for each simulated velocity range. The cuts on Nhit and $\lambda$ resulting from the optimization are indicated in Table~\ref{table1} for the three detector configurations. In order to be less dependent on the Monte Carlo statistics, an extrapolation of the background distribution of Nhit into the high Nhit region was performed. The number of atmospheric background events expected for 116 days of data taking is finally indicated in Table~\ref{table1} for each range of reconstructed velocity.

%%%%%%%
%SECTION5
\section{Results}\label{Sec_Res}

After unblinding, the remaining 85\% of the data was first compared to the simulated background after the first level selection cuts discussed before and after applying the normalization factor extracted from the 15\% data sample. A good agreement was found between data and simulation in respect to both the zenith and $\beta_{rec}$ distributions.

The number of observed events is given in Table~\ref{table1} after the final selection cuts are applied to the unblinded data sample. Only one event passes all the selection criteria and was found to lie in the range $\beta_{rec}=[0.675,0.725]$. Given the expected background of $1.3\times10^{-1}$, which requires five events for a $5\sigma$ deviation, the observation is compatible with the background-only hypothesis. Considering the observed event as background, the Feldman-Cousins 90~\% C.L. upper limits~\cite{ref:Feldman} on the upgoing magnetic monopole flux are reported in Table~\ref{table1}, for $g=g_D$, where systematic uncertainties are included.

%\subsection{Systematics}
The dominant source of systematic uncertainty is the detector efficiency for the monopole signal. The modelling of the detector efficiency depends mainly on the assumptions for the optical module angular acceptance and on the light absorption length in sea water. The detection efficiency of an optical module is determined with an uncertainty of $\pm 15\%$ for Cherenkov light from upgoing particles and the light absorption length in water is measured to $\pm 10\%$ over the whole wavelength spectrum~\cite{ref:Antares5line}. In order to estimate the effect of the detection efficiency, $18\%$ (quadratic sum of uncertainties) of hits per event were removed randomly in the Monte Carlo monopole simulation. Monopoles remaining after the selection cuts were used in the calculation, leading to a deterioration of the upper limit of 3\% for velocities above the Cherenkov threshold and 7\% for velocities below the threshold. This limit is then considered as the final one. The statistical uncertainties are negligible.

\begin{table}[!h]
\begin{footnotesize}
\caption{\footnotesize{For each velocity range of monopoles, selection cuts for all the three detector configurations are indicated, as well as the number of background events expected for the total of 116 days of live time. The number of observed events from the 85\% of unblinded data in 2008 is also reported and the Feldman-Cousins 90~\% C.L. flux limit is given, assuming that any observed events are background.}}
   \begin{center}
     \begin{tabular}{ccccccc}
     \hline
     	\rowcolor[gray]{.8} $\beta_{rec}$ & \multicolumn{3}{>{\columncolor[gray]{.8}}c}{Selection cuts  (nhit ; $\lambda$)}    &  Number of expected     & Number of  &  $90\%$ C.L. flux u. l. \\
	\cline{2-4}
     	\rowcolor[gray]{.8}      range         & 10-line  & 9-line  & 12-line                                                    & background events    &  obs. events & (cm$^{-2}$ s$^{-1} $sr$^{-1}$)\\ \hline 
	$[0.625,0.675]$ & $(27; 0.6)$   & $(28; 0.5)$      & $(36; 0.7)$   & $2.2\times10^{-2}$		&   0 & $7.5\times10^{-17}$ \\ 
	$[0.675,0.725]$ & $(34; 0.4)$   & $(35; 0.2)$      & $(47; 0.0)$   & $1.3\times10^{-1}$ 	&   1 & $8.9\times10^{-17}$ \\
	$[0.725,0.775]$ & $(43; 0.2)$   & $(57; 0.4)$      & $(53; -2.1)$  & $4.6\times10^{-2}$ 	&  0 & $4.0\times10^{-17}$ \\ 
	$[0.775,0.825]$ & $(77; 0.9)$   & $(64; 0.7)$      & $(81; 0.8)$   & $1.1\times10^{-6}$		&  0 & $2.4\times10^{-17}$ \\ 
	$[0.825,0.875]$ & $(93; 0.4)$   & $(79; 0.3)$      & $(93; 0.4)$   & $8.2\times10^{-7}$ 	&  0 & $1.8\times10^{-17}$ \\ 
	$[0.875,0.925]$ & $(118; 0.1)$ & $(99; 0.2)$      & $(85; 0.7)$   & $6.9\times10^{-7}$ 	&  0 & $1.7\times10^{-17}$ \\ 
	$[0.925,0.975]$ & $(114; 0.2)$ & $(108; 0.1)$    & $(84; 0.0)$   & $2.3\times10^{-5}$ 	&  0 & $1.6\times10^{-17}$ \\ 
	$[0.975,1.025]$ & $(85; 0.0)$   & $(110; -2.1)$   & $(92; 0.0)$   & $1.3\times10^{-2}$ 	&  0 & $1.3\times10^{-17}$ \\ \hline 

     \end{tabular}
    \label{table1}
   \end{center}
\end{footnotesize}
\end{table}

\begin{figure}[h!]
   \begin{center}
      \includegraphics[height=3.in]{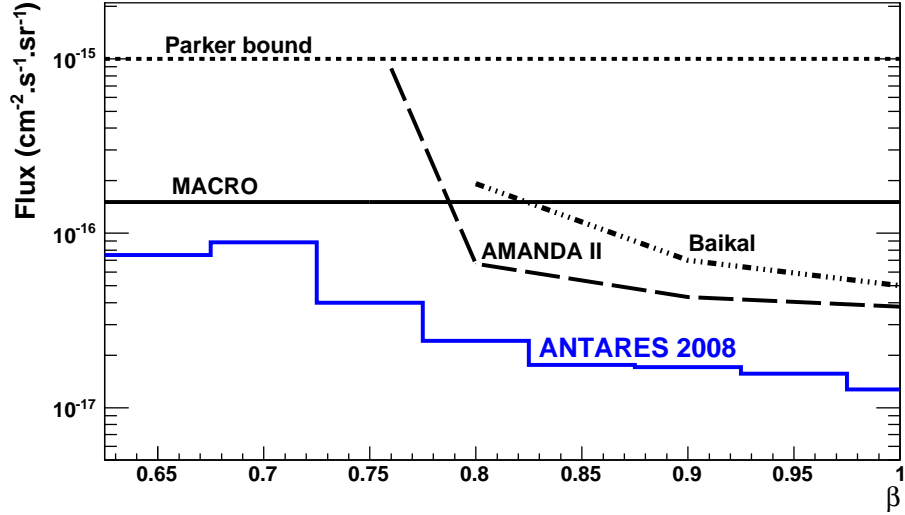}
   \end{center}
   \caption{The ANTARES 90~\% C.L. upper limit on an upgoing magnetic monopole flux for relativistic velocities $0.625\leq\beta\leq0.995$ obtained in this analysis, compared to the theoretical Parker bound~\cite{ref:Parker}, the published upper limits obtained by MACRO~\cite{ref:MACRO} for an isotropic flux of monopoles as well as the upper limits from Baikal~\cite{ref:BAIKAL} and AMANDA~\cite{ref:AMANDA} for upgoing monopoles.}
   \label{limite}
\end{figure}

%\subsection{Conclusion}
The flux limit for upgoing magnetic monopoles is shown in Fig.~\ref{limite} as a function of the monopole velocity $\beta$. The limits reported by MACRO~\cite{ref:MACRO} for an isotropic flux of monopoles, Baikal~\cite{ref:BAIKAL} and AMANDA \cite{ref:AMANDA} for upgoing monopoles are also given, as well as the theoretical Parker bound~\cite{ref:Parker}. The flux limit obtained by this analysis improves by a factor of three the upper limits on the upgoing monopole flux for velocities above the Cherenkov threshold and extends these limits to lower velocities than limits obtained by previous neutrino telescope analyses.

%%%%%%%
%SECTION5
\section{Summary}
A search for relativistic magnetic monopoles has been performed with 116 days live time of ANTARES data, yielding limits on the upgoing magnetic monopole flux above the Cherenkov threshold for $0.75\le\beta\le0.995$ ($\gamma=10$) which are more stringent than those obtained by previous experiments in this $\beta$ range. Furthermore, with a good identification of bright objects at low velocities thanks to the low light scattering in sea water, the analysis improves the upper limits below the Cherenkov threshold for $0.625\le\beta\le0.75$.

\section*{Acknowledgements} The authors acknowledge the financial support of the funding agencies: Centre National de la Recherche Scientifique (CNRS), Commissariat \`a l' \'energie atomique et aux \'energies alternatives (CEA), Agence Nationale de la Recherche (ANR), Commission Europ\'eenne (FEDER fund and Marie Curie Program), R\'egion Alsace (contrat CPER), R\'egion Provence-Alpes-C\^ote d'Azur, D\'epartement du Var and Ville de La Seyne-sur-Mer, France; Bundesministerium f\"ur Bildung und Forschung (BMBF), Germany; Istituto Nazionale di Fisica Nucleare (INFN), Italy; Stichting voor Fundamenteel Onderzoek der Materie (FOM), Nederlandse organisatie voor Wetenschappelijk Onderzoek (NWO), The Netherlands; Council of the President of the Russian Federation for young scientists and leading scientific schools supporting grants, Russia; National Authority for Scientific Research (ANCS), Romania; Ministerio de Ciencia e Innovaci\'on (MICINN), Prometeo of Generalitat Valenciana (GVA) and MultiDark, Spain. We also acknowledge the technical support of Ifremer, AIM and Foselev Marine for the sea operation and the CC- IN2P3 for the computing facilities

%\end{linenumbers}

\end{document}